\documentclass[useAMS,usedcolumn,usenatbib,usegraphicx,letters]{mn2e}

\usepackage{epsfig,graphicx,natbib}
\usepackage{./reference/mycite}
\usepackage{amssymb}
\usepackage{gensymb}
\usepackage{amsfonts}
\usepackage{amsmath}
\usepackage{color}
\usepackage{lscape,longtable}
 \usepackage{times}
 
\begin{document}

\title[ULASJ1234+0907: The Reddest Type 1 Quasar]{$ULAS J1234+0907$: The Reddest Type 1 Quasar at $z=2.5$ Revealed in the X-ray and Far Infra-red \thanks{{\it Herschel} is an ESA space observatory with science instruments provided by European-led Principal Investigator consortia and with important participation from NASA.}} 
\author[M. Banerji et al.]{ \parbox{\textwidth}
{Manda Banerji$^{1}$\thanks{E-mail: m.banerji@ucl.ac.uk}, A. C. Fabian$^{2}$ \& R. G. McMahon$^{2,3}$
}
  \vspace*{6pt} \\
$^{1}$Department of Physics \& Astronomy, University College London, Gower Street, London WC1E 6BT, UK. \\
$^{2}$Institute of Astronomy, University of Cambridge, Madingley Road, Cambridge, CB3 0HA, UK.\\
$^{3}$Kavli Institute for Cosmology, University of Cambridge, Madingley Road, Cambridge, CB3 0HA, UK.\\
} 

\maketitle

\begin{abstract} 

We present \textit{Herschel} and \textit{XMM-Newton} observations 
of ULASJ1234+0907 ($z=2.503$), the reddest broad-line Type 1 quasar 
currently known with $(i-K)_{AB}>7.1$. 
\textit{Herschel} observations indicate
that the quasar host is a hyperluminous infrared galaxy (HyLIRG) with
a total infrared luminosity of
log$_{10}$(L$_{\rm{IR}}$/L$_\odot$)=13.90$\pm$0.02.
A greybody fit gives a dust
temperature of T$_d$=60$\pm$3K assuming an emissivity index of
$\beta=1.5$, considerably higher than in submillimeter bright galaxies
observed at similar redshifts. The star formation rate is
estimated to be $>$2000M$_\odot$yr$^{-1}$ even accounting for
a significant contribution from an AGN component to the total
infrared luminosity or requiring that only the far infra-red luminosity is powered by a starburst.  \textit{XMM-Newton} observations constrain the
hard X-ray luminosity to be L$_{2-10keV}=1.3\times10^{45}$ erg/s
putting ULASJ1234+0907 among the brightest X-ray quasars known. Through very deep
optical and near infra-red imaging of the field at sub-arcsecond seeing, we demonstrate that despite
its extreme luminosity, it is highly unlikely that ULASJ1234+0907 is being lensed. We
measure a neutral hydrogen column density of
N$_H$=$9.0\times10^{21}$cm$^{-2}$ corresponding to A$_V \sim 6$.
The observed properties of
ULASJ1234+0907 - high luminosity and Eddington ratio, broad lines,
moderate column densities and significant infra-red emission from
re-processed dust - are similar to those predicted by
galaxy formation simulations for the AGN \textit{blowout} phase.
The high Eddington ratio combined with the
presence of significant amounts of dust, is expected to drive strong
outflows due to the effects of radiation pressure on dust. We conclude that ULASJ1234+0907 is 
a prototype galaxy caught at the peak epoch of galaxy formation, which is transitioning from a starburst to optical quasar
via a dusty quasar phase. 
\end{abstract}


\section{INTRODUCTION}

Since the discovery of the correlation between the mass of the
galactic bulge in galaxies and the mass of their central black holes
\citep{Magorrian:98}, understanding the link between the
formation and evolution of massive galaxies and their supermassive
black holes (SMBH; 10$^5$-10$^{10}$Msun) 
has become one of the most important problems in
both galaxy formation and extreme astrophysics. The realisation that
an active SMBH or active galactic nucleus (AGN) plays a fundamental
role in determining the final stellar mass of the galactic bulge has
wide implications, and has led to to the ubiquitous adoption of AGN
\textit{feedback} in many galaxy formation models
\citep{Croton:06}. Without AGN feedback, these
galaxy formation models cannot reproduce the observed number counts of
the most massive galaxies.

In cosmological simulations the most massive galaxies in the Universe
are assembled at high redshift through gas-rich mergers of smaller
systems \citep{Hopkins:08}. The merger is expected to induce a far
infrared (FIR) luminous starburst, which enshrouds the galaxy in dust
therefore obscuring it completely at ultraviolet (UV) and optical
wavelengths. The high gas densities in the merging starburst feed
accretion onto the central black hole, which is initially also dust
obscured. As radiative pressure on the dust grains begins to clear the
dust and gas away during the \textit{blowout} phase
\citep{diMatteo:05}, the central active galactic nucleus (AGN) is
revealed as a UV-luminous quasar. Several indirect lines of evidence
suggest that powerful starbursts and luminous quasars occur in the
same galaxies \citep{Coppin:08, Hickox:12}. However, direct
observational evidence for starburst galaxies at high
redshifts transitioning to UV-luminous quasars, has remained elusive.

In \citet{Banerji:12,Banerji:13} - B12,B13 hereafter - we used infrared
surveys to identify 14 extremely massive, luminous, dusty broad-line
quasars at $z\sim2$ that could represent galaxies caught during the AGN blowout
phase. This new population of quasars is too dusty to be present in
optical surveys like the Sloan Digital Sky Survey (SDSS) that have so
far been used to identify the most luminous quasars in the Universe.
The reddest and most intrinsically luminous quasar in our current sample
is ULASJ1234+0907 at $z=2.503$. Our spectroscopic observations in B12
detected very broad H$\alpha$ emission from this quasar signifying the
presence of an extremely massive SMBH ($\sim3\times10^{10}$M$_\odot$) and/or significant outflows that
could be broadening the H$\alpha$ line. This broad emission line also
supports the interpretation that the dust responsible for the red
colours of the quasar, originates on large scales within the quasar
host, rather than in a molecular torus. In the latter case, our view
of the quasar broad-line region would have also been blocked for most
lines of sight. The dust extinction implied by our SED fit to the
broadband colours of ULASJ1234+0907, is $A_V=6$ mags even accounting
for the effect of the large H$\alpha$ equivalent width on the $(H-K)$
colour (see B12 for details).

If ULASJ1234+0907 is
being observed in the AGN blowout phase, the next step
is to look for evidence for a massive starburst quasar host. In this
paper, we present detailed multi-wavelength observations of
this quasar. X-ray observations with \textit{XMM-Newton} allow us
to directly detect the primary accreting power source while far infrared and submillimeter observations with
\textit{Herschel} and SCUBA-2 are used to place the first constraints on the
quasar host galaxy. Throughout this paper we assume a flat
$\Lambda$CDM cosmology. All magnitudes are on the AB system with
conversion from the Vega system using zeropoint offsets for
UKIDSS and \textit{WISE} photometry from \citet{Hewett:06}
and \citet{Cutri:12}

\section{OBSERVATIONS}

\label{sec:data}

\subsection{Optical and Near Infra-red Imaging}

\label{sec:opt}

We conducted $i$-band imaging of the quasar using the Wide-Field
Camera on the 2.5m ISAAC Newton Telescope in April and May 2008
The total exposure time was 85 minutes.
The six individual exposures were reduced, stacked and 
photometrically calibrated onto the SDSS 
$AB$ system following \citet{Gonzalez:11}.
ULASJ1234+0907 was undetected in the stacked $i$-band
image (Figure \ref{fig:cutouts}) 
with a 5$\sigma$ magnitude limit of 25.15
giving an $(i-K)$ colour of $>$7.1. ULASJ1234+0907 is therefore among the reddest extragalactic
sources known. Figure 2 of B12 shows the
$(i-K)$ colours of various well-known extremely red objects.
The reddest quasar from that figure:
PKSJ0132 which was the reddest quasar previously known
\citep{Gregg:02} has an $(i-K)$=4.2, almost 3
mags bluer than ULASJ1234+0907.

We conducted $K_S$ band observations of
ULASJ1234+0907 using the ISAAC camera on the Very Large Telescope. Data was taken in March 2013 under photometric
conditions and in very good seeing of $<$0.5$^{\prime \prime}$ 
The observations were taken using 15 dithered exposures of
60s and reduced using standard ESO
tools provided as part of the \textit{gasgano} package. 
The resulting $K$-band
image shown in Figure \ref{fig:cutouts} goes down to a 10$\sigma$ depth of 
23.2.

\subsection{Far Infrared and Submillimeter Photometry: \textit{Herschel} \& SCUBA-2}

\label{sec:herschel}

\begin{figure}
\begin{center}
\includegraphics[scale=0.5,angle=0]{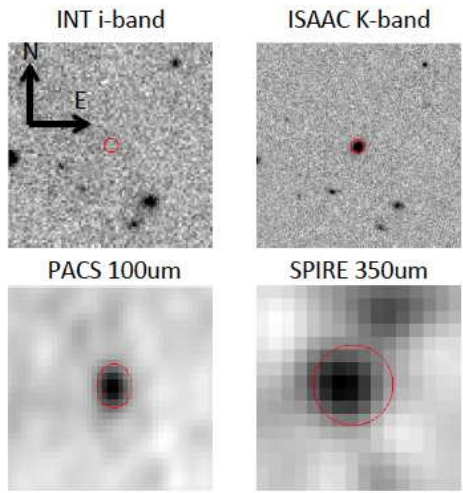}
\caption{Multi-wavelength images of ULASJ1234+0907 taken with various instruments. The INT $i$-band and ISAAC $K_S$-band images are 30$\times$30$^{\prime \prime}$ while the \textit{Herschel} cut-outs are 2$\times$2$^{\prime}$ in size. In all cases, the direction of north and east are as specified in the $i$-band image. The quasar position is circled in all images with the size of the circle roughly corresponding to the beam size at that particular wavelength.}
\label{fig:cutouts}
\end{center}
\end{figure}

\textit{Herschel} \citep{Pilbratt:10} Director's Discretionary Time 
observations were obtained
in December 2012 with PACS 
\citep{Poglitsch:10}.
Data was taken in mini scanmap mode using a 
scan speed of 20$^{\prime \prime}$/s and a scan leg length of
3$^{\prime}$ with exposure times in the range of 444 to 895s
at 70$\mu$m, 100$\mu$m and 160$\mu$m.
PACS maps 
were produced 
with the default pixel
scale of 1.4$^{\prime \prime}$, 1.7$^{\prime \prime}$ and
2.8$^{\prime \prime}$ at 70, 100 and 160$\mu$m respectively
using \textit{Herschel} Level 1 data products and the map-making 
software \textit{Scanamorphos} \citep{Roussel:12}. 
Fluxes were measured by performing annular aperture photometry at the
quasar position using apertures of 5.5$^{\prime
  \prime}$, 5.6$^{\prime \prime}$ and 10.5$^{\prime \prime}$ at 70,
100 and 160$\mu$m respectively. The background was measured using
apertures of [20,25]$^{\prime \prime}$ at 70 and 100$\mu$m and
[24,28]$^{\prime \prime}$ at 160$\mu$m.  Appropriate aperture
corrections were applied to these fluxes \citep{Poglitsch:10}. Statistical errors on the fluxes
were determined by taking the standard deviation of aperture fluxes
calculated in blank fields on the combined maps. Systematic flux
calibration errors of 5\% were added in quadrature to these. The final
fluxes and errors are presented in Table \ref{tab:flux} and the PACS
cutout at 100$\mu$m  is shown in Figure \ref{fig:cutouts}.

\begin{table}
\begin{center}
\caption{Summary of Multi-wavelength Observations of ULASJ1234+0907 at $z=2.503$}
\label{tab:flux}
\begin{tabular}{lcc}
\hline \hline
Instrument \& Band & Flux Density  \\
\hline
\hline
INT $i$ & $<$0.316 $\mu$Jy (5$\sigma$)  \\
WFCAM $Y$ & $<$0.435 $\mu$Jy (5$\sigma$)  \\
WFCAM $J$ & 4.06$\pm$1.10 $\mu$Jy  \\
WFCAM $H$ & 25.9$\pm$4.9 $\mu$Jy  \\
WFCAM $K$ & 219.4$\pm$6.1$\mu$Jy \\
WISE 3.4$\mu$m & 0.48$\pm$0.02 mJy  \\
WISE 4.6$\mu$m & 1.14$\pm$0.04 mJy \\
WISE 12$\mu$m & 6.27$\pm$0.23 mJy \\
WISE 22$\mu$m & 10.1$\pm$1.6 mJy \\
\textit{Herschel}-PACS 70$\mu$m & 41$\pm$4 mJy \\
\textit{Herschel}-PACS 100$\mu$m & 58$\pm$6 mJy \\
\textit{Herschel}-PACS 160$\mu$m & 83$\pm$8 mJy  \\
\textit{Herschel}-SPIRE 250$\mu$m & 57$\pm$9 mJy  \\
\textit{Herschel}-SPIRE 350$\mu$m & 59$\pm$9 mJy \\
\textit{Herschel}-SPIRE 500$\mu$m & 50$\pm$12 mJy \\
SCUBA-2 850$\mu$m & $<$12 mJy (3$\sigma$)  \\
\hline
\end{tabular}
\end{center}
\end{table}

\textit{Herschel}-SPIRE data at 250, 350 and 500$\mu$m, is available
as part of the \textit{Herschel} Virgo Cluster Survey (HeViCS;
\citealt{Davies:10}) in the HeViCS V3 field. 
We used the Level 1 processed
data from the \textit{Herschel} Science Archive.
Timeline
fitting was done on each of the 4 orthogonal scans within HIPE using
sourceExtractorTimeline. 
In several cases, the timeline
fitting produced fluxes for one of the scans that was inconsistent
with the remaining scans at the level of 20\%. We therefore
rejected the most discrepant scan from the set of four observations
and took the final SPIRE flux to be the median value of the remaining
scans. Colour corrections of 0.912, 0.916 and 0.901 at 250, 350 and 500$\mu$m were applied, appropriate for
a typical dusty source with spectral index $\alpha=3$. The final fluxes are presented in Table \ref{tab:flux}. Error estimates from the timeline fitting were very
similar from scan to scan: $\sim$8mJy, 8mJy and 10mJy at 250, 350 and
500$\mu$m respectively. 
Our final errors include the calibration uncertainty 
assumed to be 7\% \citep{Bendo:13} in addition to
the PSF fitting errors. 

SCUBA-2 service time observations 
(s12au01; PI:Banerji) were taken in April 2012. 
The source was
observed in \textit{Daisy} mode for 20 minutes and the data was reduced using
the SubMillimeter User Reduction
Facility (SMURF) iterative map maker software to construct an 850$\mu$m flux
density map of the quasar using the default configuration parameters
for blank fields. The reduced image has a mean RMS of
4mJy. The quasar was undetected in this 850$\mu$m image leading to a
3$\sigma$ upper limit of S$_{850}<$12mJy for ULASJ1234+0907. 

\subsection{X-ray Observations}

X-ray data for ULASJ1234+0907 was obtained using the the European Photon Imaging Camera (EPIC) on the \textit{XMM-Newton} satellite \citep{Jansen:01}. The total exposure time is 42ks for the pn detectors and 52ks for the MOS detectors. Thin filters were used. The total pn source count is 470. A spectrum was extracted in a 21.6$^{\prime \prime}$ aperture around the target. The background was measured from the same chip at a radius of 1.2$^{\prime}$ from the source. The X-ray spectrum from the pn detector can be seen in Figure \ref{fig:X-ray}. The MOS detectors are less sensitive but the spectrum obtained from these is consistent with Figure \ref{fig:X-ray}. 

\section{ANALYSIS}

\label{sec:analysis}

\subsection{SED Fitting and Star Formation in ULASJ1234+0907}

We fit a spectral energy distribution (SED) model to the observed photometry of ULASJ1234+0907 summarised in Table \ref{tab:flux}. Fitting the SED in the far infrared requires us to disentangle the contribution of the AGN and the starburst to the total infrared luminosity. We adopt two approaches for the SED fitting. Firstly, we use the publicly available SED fitting code CIGALE \citep{Noll:09}, which allows us to fit for both the AGN and starburst components at infrared wavelengths. In this code, the fractional contribution of the AGN to the total infrared luminosity is left as a free parameter in the fitting. The code uses the semi-empirical models of \citet{Dale:02} to model the dust emission at infrared wavelengths. AGN templates can additionally be incorporated and we adopt 32 AGN templates from the library of \citet{Fritz:06}, encompassing a range of torus opening angles, density parameters, optical depths and outer and inner radii of the dust clouds. We note that the inclusion of other AGN templates available within CIGALE, does not significantly change our results. 

Our second approach is simpler with fewer free parameters and allows direct comparison with greybody fits to the SEDs of other high redshift galaxy populations. We use the code developed by \citet{Casey:12} to fit a single temperature greybody and a power-law of the form $S_\lambda \propto \lambda^{\alpha}$, to model the mid infrared emission at rest-frame wavelengths $\gtrsim$3$\mu$m. The power-law approximates the contribution from an AGN-heated warm dust component. 

Using the CIGALE fit we find that the total infrared luminosity is log$_{10}$(L$_{\rm{TIR}}$)=13.90$\pm$0.02 with the AGN contribution to this IR luminosity fit to be 62$\pm$4\%. The starburst luminosity is therefore log$_{10}$(L$_{\rm{SB}}$)=13.5$\pm$0.1 which translates to a star formation rate of $\sim$4500$\pm{900}$ M$_\odot$yr$^{-1}$ using the \citet{Kennicutt:12} relation: $\rm{SFR}=3.89 \times 10^{-44} \times L_{\rm{SB}} (\rm{erg s}^{-1})$
Despite the AGN dominating the infrared dust emission, the SED fitting indicates that this AGN cannot account for all of the infrared emission and the host galaxy of ULASJ1234+0907 must also be forming stars at a prodigious rate. With an inferred starburst luminosity of $>10^{13}L_\odot$, it can be classified as a hyperluminous infrared galaxy or HyLIRG.

In the case of the single greybody fit, we fix the dust emissivity index to $\beta=1.5$ \citep{Priddey:03} due to a lack of photometric data over the Rayleigh-Jeans tail of the SED. The free parameters are the power-law slope, $\alpha$ in the MIR as well as the dust temperature. Although the AGN is expected to dominate the total infra-red luminosity, the far infra-red luminosity between 40-300$\mu$m can safely be attributed to a starburst \citep{RR:00}. We integrate our best-fit single greybody SED between 40 and 300$\mu$m and use this as a conservative estimate of the total amount of star formation. The infrared luminosity between 40-300$\mu$m, L$_{40-300\mu\rm{m}}=1.3\times10^{13}$L$_\odot$ corresponding to a star formation rate of $\sim$2000$\pm{500}$ M$_\odot$yr$^{-1}$. The mid infra-red power law slope is relatively shallow: $\alpha$=1.03$\pm$0.03, which is consistent with the presence of a significant warm dust component from AGN heated dust. The best-fit cold dust temperature is also significantly higher than seen in most submillimeter galaxies: T$_d$=60$\pm$3K.  The corresponding dust-mass is log$_{10}$(M$_{\rm{dust}}$/M$_\odot$)=$8.94\pm0.08$. 

\begin{figure}
\begin{center}
\includegraphics[scale=0.45,angle=0]{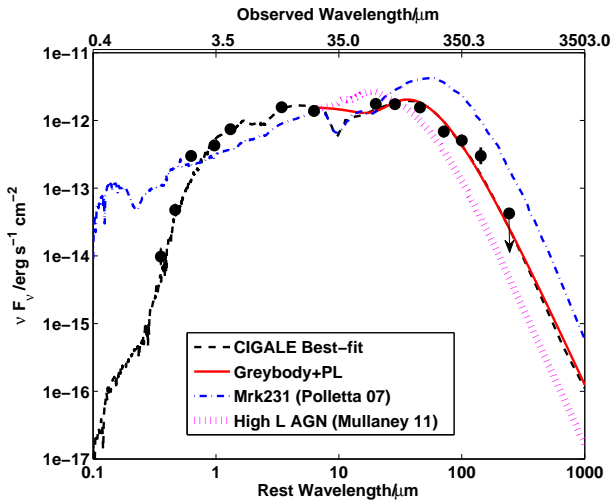}
\caption{Best-fit SED to the observed photometry of ULASJ1234+0907 at $z=2.503$ in units of $\nu F_\nu$. The dashed line is the best-fit CIGALE SED which includes both AGN and starburst components while the solid line denotes the simple greybody plus mid-IR power-law fit to the data. We also show the SEDs of Mrk231 from \citet{Polletta:07} (dot-dashed line) and the high luminosity AGN from \citet{Mullaney:11} (dotted line). All SEDs have been scaled to match the rest-frame flux density of ULASJ1234+0907 at $\sim$6$\mu$m which corresponds to the observed-frame \textit{WISE} 22$\mu$m band.}
\label{fig:sed}
\end{center}
\end{figure} 

In Figure \ref{fig:sed}, we also show the SED of the archetypal starburst/AGN composite galaxy Mrk231 in the local Universe which has been scaled to have the same flux density as ULASJ1234+0907 at a rest-frame wavelength of 6$\mu$m. Finally we plot the composite far infrared SED of the high luminosity (log($L_{2-10\rm{keV}})>42.9$) X-ray AGN from \citet{Mullaney:11} again scaled to ULASJ1234+0907 at a rest-frame wavelength of 6$\mu$m. ULASJ1234+0907 has a higher mid to far infra-red flux ratio than Mrk231. The pure AGN SED on the other hand, drops rapidly at rest-frame wavelengths of $\gtrsim$40$\mu$m whereas the SED of ULASJ1234+0907 extends to longer wavelengths and only begins to drop off at $\sim$60$\mu$m. This excess cool dust emission relative to pure AGN, could again be indicative of a starburst component. Further evidence for star formation is provided by the fact that the observed 500$\mu$m photometric data-point is not particularly well-fit by the models and lies above the model SEDs. At the redshift of this quasar, the [CII] cooling line is in the SPIRE 500$\mu$m band and can increase the broadband flux by 20-40\% \citep{Smail:11}. 

Our observations cannot directly rule out the presence of AGN heated cool dust distributed at larger radii from the central black-hole than the hotter dust responsible for the mid infrared emission. However, many FIR luminous quasars at high redshift have been demonstrated to contain significant reservoirs of molecular gas, which is unambiguous evidence for star formation \citep{Wang:10}. Furthermore, we demonstrate below using X-ray observations of the quasar, that the AGN bolometric luminosity inferred independently from the X-ray, is significantly lower than the total infra-red luminosity measured for this object. This once again hints that some additional source of heating apart from the AGN, may be responsible for the reprocessed far infra-red dust-emission. Regardless of the origin of the cool dust emission in the far infrared, we conclude that ULASJ1234+0907 is among the most far infrared luminous broad-line quasars known. 


\subsection{X-Ray Spectrum}

\begin{figure}
\begin{center}
\includegraphics[scale=0.3, angle=270]{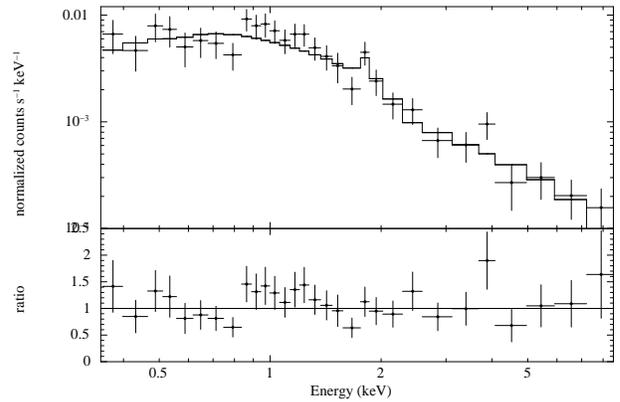}
\caption{X-ray spectrum of ULASJ1234+0907 in the observed-frame together with the best-fit absorbed power-law model to this spectrum. The spectrum covers observed-frame energies of 0.4-8 keV corresponding to rest-frame energies of 1.4-28 keV. }
\label{fig:X-ray}
\end{center}
\end{figure}

In Figure \ref{fig:X-ray}, we plot the X-ray spectrum of ULASJ1234+0907 obtained using \textit{XMM-Newton} together with the best-fit model spectrum fit using XSPEC. The model includes Galactic absorption plus intrinsic absorption (at $z=2.5$) acting on a power law continuum with a narrow emission line at 6.4 keV to account for fluorescent Fe K$\alpha$. The total flux in the hard X-ray band covering $2-10$ keV (observed frame) is $2.6\times10^{-14}$ erg s$^{-1}$cm$^{-2}$ from the model fit. The (source rest-frame) hard X-ray luminosity is therefore L$_{2-10keV}=1.3\times10^{45}$ erg/s in our adopted cosmology. The best-fit photon index is 1.68$\pm^{0.25}_{0.22}$ and the inferred hydrogen column density is N$_H=(9.0\pm^{1.0}_{0.8})\times10^{21}$cm$^{-2}$. This column density agrees with the extinction of Av=6 mags calculated from the near infrared broad-band colours in B12 assuming the dust properties are similar to our own Milky Way. The X-ray observations therefore indicate that ULASJ1234+0907 is also among the most X-ray luminous quasars known with a hard X-ray luminosity that is almost two orders of magnitude greater than Mrk231 and similar to the most powerful quasars in the Universe.  

In the most powerful quasars accreting at close to the Eddington limit, the bolometric correction from the hard X-ray band is expected to be $\sim$50-100 \citep{Elvis:94}. This implies that the AGN bolometric luminosity of ULASJ1234+0907 is $\sim1.3\times10^{47}$ erg/s. Even assuming that all of this luminosity is seen as reprocessed dust emission in the far infrared, the shortfall in the total infrared luminosity calculated from the Herschel data, is still $\sim1.7\times10^{47}$ erg/s. If this \textit{missing} infrared luminosity is to come from star formation, we now obtain an even more extreme star formation rate in the quasar host galaxy of $\sim$6800 M$_\odot$/yr. 
Conversely, we can consider the bolometric luminosity of ULASJ1234+0907 derived from the de-reddened UV luminosity: L$_{\rm{bol}}=1.9\times10^{48}$ erg/s (B12). Based on this estimate of the bolometric luminosity and assuming a bolometric correction of $\sim$100, the X-ray luminosity is predicted to be $1.9\times10^{46}$ erg/s, more than an order of magnitude larger than what we measure. These discrepancies could indicate that the source is potentially X-ray weak. 


The census of known AGN shows a remarkable lack of luminous, high Eddington ratio quasars that also have high columns of dust \citep{Raimundo:10}. These conditions are expected to drive strong outflows due to the effects of radiation pressure on the dust \citep{Fabian:12}. The lack of very luminous, dusty quasars is perhaps not surprising given that most rapidly accreting quasars with high Eddington ratios have predominantly been selected at UV-optical wavelengths. In ULASJ1234+0907, the A$_V$=6 mags at $z=2.5$ corresponds to 17.5 mags of extinction in the observed frame optical $i$-band. In other words, the quasar flux is suppressed by a factor of $\sim$10$^{7}$ at optical wavelengths due to absorption by dust making it invisible in even the deepest optical surveys coming up within the next decade e.g. LSST. 

Highly reddened Type 1 quasars like ULASJ1234+0907 that are accreting at close to the Eddington limit, are distinct from the well-studied obscured TypeII AGN population. The FeK emission at rest-frame 6.4keV (observed-frame 1.8 keV) in the quasar spectrum in Figure \ref{fig:X-ray} is only marginally detected. The X-ray spectrum of ULASJ1234+0907 is therefore markedly different from typical Compton thick sources like NGC1068 where the FeK$\alpha$ line is much more prominent in the reflection dominated spectrum, and the photon index is considerably less steep. \citep{Iwasawa:97}.  

 \subsection{Evidence against lensing}

We have demonstrated that ULASJ1234+0907 is among the brightest and reddest broad-line quasars currently known with extremely high luminosities measured at both far infra-red and X-ray wavelengths. Given these extreme luminosities, we should consider the possibility of the quasar being lensed. The strongest evidence against lensing is provided by our deep $i$ and $K$-band observations (Section \ref{sec:data}). The optical image goes down to a magnitude limit of $i_{AB}<25.15$. Even a high-redshift galaxy lens at $z > 1$ should have been easily visible at these depths. The quasar is undetected in the $i$-band and there are no other galaxies present within a 7$^{\prime \prime}$ radius of the quasar position in this $i$-band image so there is no evidence to support the presence of a galaxy lens. 

The $K$-band data from ISAAC was taken in $\sim$0.5$^{\prime \prime}$ seeing which would allow us to easily resolve multiple emission sources associated with lensed images of the quasar, should they have been present. From Figure \ref{fig:cutouts}, we see that there is no evidence for multiple images in this sub-arcsecond seeing $K$-band data. The $K$-band emission from the quasar is unresolved and consistent with the size of the PSF. The lensing cross-section is therefore extremely small and the probability of lensing very low. 

\section{DISCUSSION \& CONCLUSIONS}

The above suite of multi-wavelength observations has allowed us to build up a comprehensive picture of the properties of ULASJ1234+0907, the reddest broad-line quasar currently known. With an X-ray luminosity of L$_{(2-10)keV}$=1.3$\times$10$^{45}$ erg/s and a total infrared luminosity of L$_{\rm{TIR}}$=3.1$\times$10$^{47}$ erg/s, ULASJ1234+0907 is among the most luminous quasars known at both these wavelengths. Through sub-arcsecond, deep imaging of the quasar in the $i$ and $K$-bands, we demonstrate that no galaxy lens is apparent in the optical and that the quasar emission is unresolved in the $K$-band. Lensing is therefore unlikely to be responsible for the extreme luminosities of this source. Even accounting for a large (62\%) contribution of the AGN to the total infrared luminosity or assuming that only the FIR luminosity between 40 and 300$\mu$m is powered by a starburst, the estimated SFR in the host galaxy of this quasar is $>$2000M$_\odot$yr$^{-1}$. Measuring the AGN bolometric luminosity directly from the hard X-ray luminosity (assuming a bolometric correction of 100) and therefore requiring the rest of the total infra-red emission to be powered by a starburst, gives an even more extreme SFR of $\sim$6800M$_\odot$yr$^{-1}$. The column density inferred from the X-ray spectrum is N$_H=9.0\times10^{21}$cm$^{-2}$ corresponding to A$_V$=6. 

Cosmological simulations, predict the existence of a high-Eddington ratio, luminous, reddened quasar phase which marks the transition of massive starburst galaxies to UV luminous quasars. This phase is distinct from the well-known Type II obscured AGN \citep{Hopkins:08}. During this phase, the bolometric output of both the starburst and AGN are expected to be at their peak. ULASJ1234+0907 has all the observed properties expected, were it to be detected during this transition phase - high luminosity and Eddington ratio, significant dust-reddening, broad emission lines and re-processed dust emission at far infra-red wavelengths. This quasar is an order of magnitude more luminous than the well-studied Submillimeter Galaxies (SMGs) and Dust Obscured Galaxies  which have typically been selected over much smaller areas with \textit{Spitzer} and/or SCUBA imaging. The black-hole mass is also considerably larger than measured in SMGs (B12). In terms of luminosity and space density, ULASJ1234+0907 is better matched to the new population of HyLIRGs discovered using the \textit{WISE}- All Sky Survey \citep{Eisenhardt:12} although drawing evolutionary links between the \textit{WISE} HyLIRGs and dust-reddened Type 1 quasars, would require detailed consideration of the relative lifetimes of these two phases, which is beyond the scope of this paper. We emphasise that reddened Type 1 quasars like ULASJ1234+0907 are distinct from the more highly obscured and less luminous Type II AGN where the dust extinction can often be explained by orientation effects. We have provided some evidence that the dust extinction in ULASJ1234+0907 arises on larger scales in the quasar host galaxy thus allowing the broad-line region to be viewed in the near infra-red. This dust is likely heated by both the AGN and a powerful starburst as would be expected during blowout. Although our observations do not directly detect AGN driven outflows, the combination of a supermassive black-hole accreting at close to the Eddington limit coupled with the high column density, is expected to drive strong outflows due to the effects of radiation pressure on dust.

The reddened quasar phase is unique in allowing us to study both the central accreting power-source (through X-ray and NIR spectroscopy) and the starburst host galaxy (through FIR photometry) in massive galaxies observed at the peak epoch of galaxy and black-hole formation. We are assembling multi-wavelength observations of much larger samples of these highly reddened Type 1 quasars which will inform theories of massive galaxy formation and be ideal testbeds for studying AGN feedback.

\section*{Acknowledgements}

MB would like to thank Paul Hewett for many useful discussions. We thank Matt Auger for discussions on lensed quasars and Maud Galametz for help with the \textit{Herschel} data reduction. 
Based on observations made with XMM-Newton, an ESA science mission with instruments and contributions directly funded by ESA Member States and NASA and the  ESO Telescopes at the La Silla Paranal Observatory under programme ID: 290.A-5062.  

\footnotesize{
\bibliography{}
}

\end{document}